\documentclass[twocolumn,floatfix,aps,prl,superscriptaddress]{revtex4-2}

\usepackage{amsmath,amssymb,mathtools}
\usepackage{graphicx}
\graphicspath{{../}{./}}
\usepackage[colorlinks=true,linkcolor=blue,citecolor=blue,urlcolor=blue]{hyperref}

\newtheorem{proposition}{Proposition}

\newtheorem{corollary}{Corollary}

\newcommand{\Tr}{\operatorname{Tr}}
\newcommand{\E}{\mathbb E}
\newcommand{\I}{\mathbb I}
\newcommand{\Swap}{\mathbb S}
\newcommand{\cE}{\mathcal E}
\newcommand{\cH}{\mathcal H}
\newcommand{\cL}{\mathcal L}
\newcommand{\Haar}{\mathrm {Haar}}
\newcommand{\ket}[1]{|#1\rangle}
\newcommand{\bra}[1]{\langle #1|}
\newcommand{\norm}[1]{\left\|#1\right\|}

\usepackage{xcolor} 
\usepackage[normalem]{ulem}   

\DeclarePairedDelimiter{\abs}{\lvert}{\rvert}

\begin{document}

\title{When Randomness Hides Quantum Simulation Errors}

\author{Yue Cao}
\affiliation{QICI Quantum Information and Computation Initiative, School of Computing and Data Science,
The University of Hong Kong, Pokfulam Road, Hong Kong SAR, China}
\author{Tianfeng Feng}
\affiliation{QICI Quantum Information and Computation Initiative, School of Computing and Data Science,
The University of Hong Kong, Pokfulam Road, Hong Kong SAR, China}
\author{Jonathan Allcock}
\email{jonallcock@tencent.com}
\affiliation{Tencent Quantum Laboratory, Shenzhen, China}
\author{Chenfeng Cao}
\affiliation{HK Institute of Quantum Science and Technology, The University of Hong Kong, Pokfulam Road, Hong Kong SAR, China}
\author{Qi Zhao}
\email{zhaoqi@cs.hku.hk}
\affiliation{QICI Quantum Information and Computation Initiative, School of Computing and Data Science,
The University of Hong Kong, Pokfulam Road, Hong Kong SAR, China}

\date{\today}

\begin{abstract}

Trotter formulas approximate quantum dynamics by discrete steps, but errors in measured observables can be much smaller than worst-case estimates. 
We investigate how this suppression depends on the degree of randomness in the input states. We show that the mean absolute observable-error bound for state 2-designs can be exponentially smaller in qubit number than the tight bound for state 1-designs. 
To characterize this error transition from 1-design to 2-design, we use the
excess frame potential to quantify the ensemble's second-moment distance.
We quantify departures from these random-state statistics through the frame potential, a measure of pairwise state overlaps, and identify the accessible observables most sensitive to the resulting deviations in mean-square error. 
Extending the analysis to a Hamiltonian trajectory whose frame potential decays as $1/T$ before saturation, 
we also obtain a
sufficient $p$th-order Trotter number
$\mathcal O(T^{1+3/(4p)})$ for evolution time $T$.
These findings connect input-state statistics to the visibility of Trotter errors, giving a theoretical guarantee and foundation for practical quantum simulation. 
\end{abstract}

\maketitle

\textit{Introduction.---}
Hamiltonian simulation is a central application of quantum computers
\cite{feynman1982simulating,abrams1997simulation, childs2021theory,wang2026advances}.  Its applications range from
spin and strongly correlated lattice models to quantum chemistry and quantum
field theory
\cite{somma2002simulating,wecker2015strongly,wecker2014gate,
babbush2015chemical,jordan2012quantum}. Linear combination of unitaries and quantum signal processing provide efficient approaches to Hamiltonian simulation \cite{LCUPhysRevLett.114.090502, low2017optimal, low2019hamiltonian}, while quantum singular value transformation extends these ideas to a general framework for quantum matrix transformations \cite{gilyen2019quantum}.  Among digital simulation methods,
Suzuki--Trotter product formulas are particularly direct: they approximate a
target evolution by alternating evolutions under experimentally available
Hamiltonian terms \cite{suzuki1991general,childs2021theory}.  For a $p$-th order
formula, $U_r(t)=[S_p(t/r)]^r$ approximates
$U_0(t)=e^{-iHt}$. Here $S_p(\delta t)$ is one product-formula step built from exponentials
of the Hamiltonian terms, approximating evolution over $\delta t=t/r$ with
operator-norm error $\mathcal O(\delta t^{p+1})$.
Standard commutator-scaling bounds give
$\norm{U_0(t)-U_r(t)}
=\mathcal O(\widetilde\alpha_{\rm comm}t^{p+1}/r^p)$
\cite{childs2021theory}, where $\|\cdot\|$ denotes the operator norm and
$\widetilde\alpha_{\rm comm}$ is the standard nested-commutator scale.  This is
a worst-case-state guarantee.  The Trotter number $r$ therefore converts
an error estimate directly into circuit depth.

This worst-case guarantee is intentionally independent of both the input state
and the measured quantity.  In practice, however, experiments are sensitive to the error in a given observable $M$ -- given by $\bra{\psi}[U_0(t)^\dagger M U_0(t)-U_r^\dagger(t) M U_r(t)]\ket{\psi}$ -- 
rather than the
operator-norm distance between two propagators. The error  of observable simulation and its dependence on the measured quantity have been studied in digital and analog quantum simulation~\cite{heyl2019quantum,poggi2020quantifying,li2024principal,yu2024observable,feng2025trotterization}.
While random-input analyses can replace
the operator norm by a normalized Frobenius scale \cite{zhao2022hamiltonian,Chen_2024}-- effectively shifting from a worst-case to an average-case error analysis -- incorporating observable- and state-specific details can sharpen
product-formula guarantees even further
\cite{li2024principal,yu2024observable,zhao2025entanglement,
feng2025trotterization,rao2026stabilitydigitalanalogquantum}.  Because the required circuit depth in Trotter-based Hamiltonian simulation depends directly on the chosen error metric, identifying the actual state-dependent error suppression is a practical matter of physical resources, not merely a formal choice of norm.

To quantify this state-dependent suppression, the natural operational target is the expected absolute observable error, $\mathbb{E}_{\psi\sim\mathcal{E}} \abs{\bra{\psi} B(M)\ket{\psi}}$. Here $B(M) = U_0^\dagger M U_0 - U^\dagger M U $ is the Heisenberg-picture error observable for an implemented evolution $U$ (such as the Trotter approximation $U_r$) of a desired ideal evolution $U_0$, and the expectation is over states $\psi$ drawn from an ensemble $\mathcal{E}$.  A state
1-design -- an ensemble matching the first moment of the Haar ensemble -- already provides an upper bound on this expected error. However,  a state
2-design, which additionally captures the variance, can substantially improve its dimension scaling.  As we show in
Proposition~\ref{prop:design-simulation-gap}, a 1-design bounds the error by $\norm{B(M)}_1/d$, whereas a 2-design tightens this bound to $\norm{B(M)}_F/\sqrt{d+1}$, where $B(M)$ is the error observable and $d$ is the Hilbert-space dimension. Consequently,  the 2-design bound is smaller by a
factor of order $\sqrt d$ when the error observable's trace norm is of order
$d$ times its normalized Frobenius norm. 
This dimension-dependent gain highlights that random-state benchmarks can reveal or conceal a Trotter error
depending on the sampled ensemble.  State designs and frame potentials provide
the natural language for this distinction
\cite{ambainis2007quantum, scott2006tight,gross2007evenly,dankert2009,
harrow2009,brandao2016local}. Physically, a frame potential measures the average pairwise overlap between states drawn from the ensemble, capturing how uniformly they are distributed across the Hilbert space. The excess second frame potential  -- the gap between this measured overlap and the minimum baseline achieved by true randomness -- precisely quantifies the
ensemble's second-moment departure from the Haar value.

Here, we characterize how the excess frame potential governs observable Trotter errors.   To bound
the mean absolute error for pure-state 1-design inputs, we decompose the mean
square into a Haar baseline and an ensemble-dependent residual:  $\mathcal{V}_{\cE}(M) = \mathcal{V}_{\mathrm{Haar}}(M) + \text{residual}$. The excess
second frame potential fixes the total weight of the residual. The ensemble's stable rank -- a measure of its effective dimensionality -- then quantifies how sharply that error is concentrated into specific operator directions, and the Heisenberg error map filters which of those components actually manifest in measured errors.
 By projecting this residual onto the space of accessible measurements, we can identify both the maximum possible deviation from the Haar baseline and the specific physical observable that produces it.  We demonstrate this behavior explicitly: Product-Pauli (a 1-design) and Haar-random inputs clearly illustrate the gap between high-error and low-error ensembles. Furthermore, applying random quantum circuits to product-Pauli input states systematically erases this residual error, demonstrating that the deviation vanishes entirely as the ensemble converges to a 2-design.

 
The frame-potential perspective naturally extends to temporal ensembles generated along a single Hamiltonian trajectory. Such ensembles have been studied in quantum equilibration~\cite{linden2009quantum} and Hilbert-space ergodicity~\cite{pilatowsky2023complete,mark2024maximum}, with their second-moment structure further characterized through temporal frame potentials by Hou \emph{et al.}~\cite{hou2026state}. Building on this connection, we extend our analysis to accumulated Trotter errors along a single Hamiltonian trajectory, without requiring the temporal ensemble to form a state design. By decomposing the accumulated error into single-step state-vector errors along successive exact states, we relate the mean absolute endpoint error, averaged over initial time offsets $s\in[0,\delta t]$ with $\delta t=T/r$, to the temporal frame potential. As established in Proposition~\ref{thm:trajectory-trotter-scaling}, in a pre-saturation regime where $F_2(T)=\mathcal O(T^{-1})$, this yields a sufficient $p$th-order Trotter step count $r=\mathcal O(T^{1+3/(4p)})$ at fixed accuracy.

\textit{The 1-design--2-design Trotter-error gap.---}
Let $\cH=(\mathbb C^2)^{\otimes n}$ with $n\ge2$ and $d=2^n$, and let
$\cE=\{p_a,\ket{\psi_a}\}_a$ be a pure-state ensemble with
$\rho_a=\ket{\psi_a}\bra{\psi_a}$.  We denote its $k$-th moment by
\begin{equation}
M_k(\cE):=\E_{\psi\sim\cE}\rho_\psi^{\otimes k}
\end{equation}
for $k=1,2$.
The ensemble is a state 1-design if $M_1(\cE)=\I/d$, and a state 2-design if
$M_2(\cE)=\frac{\I+\Swap}{d(d+1)}$, where $\Swap$ is the swap operator on $\cH\otimes \cH$. Note that every state 2-design is also a state 1-design.
Let $U_0$ be the ideal evolution and $U$ the implemented evolution.  
Throughout our application,
$U$ is a Trotter product formula.  For a Hermitian observable $M$,
we define the Heisenberg-picture error,
$
B(M):=U_0^\dagger M U_0-U^\dagger M U.
$
This operator is Hermitian and traceless. 
For a given input state, the observable error is
\begin{align}
    X_\psi(M)&:=\bra{\psi}B(M)\ket{\psi}.
\end{align}
Our primary target is to bound the mean absolute error, $\E_{\cE}\abs{X_\psi(M)}$. We evaluate this via the mean square, $\mathcal V_{\cE}(M):=\E_{\psi\sim\cE}\abs{X_\psi(M)}^2$, using the inequality $\E_{\cE}\abs{X_\psi(M)}\le\sqrt{\mathcal V_{\cE}(M)}$. To quantify these bounds, we use the normalized Frobenius norm $\norm{X}_F:=\sqrt{\Tr(X^\dagger X)/d}$, alongside the standard spectral norm $\norm{\cdot}$ and trace norm $\norm{\cdot}_1$. Besides, we fix the Hermitian Pauli basis such that $d^{-1}\Tr(P_\mu P_\nu)=\delta_{\mu\nu}$ with $P_0=\I$.

\begin{figure}[t]
\includegraphics[width=1\columnwidth]{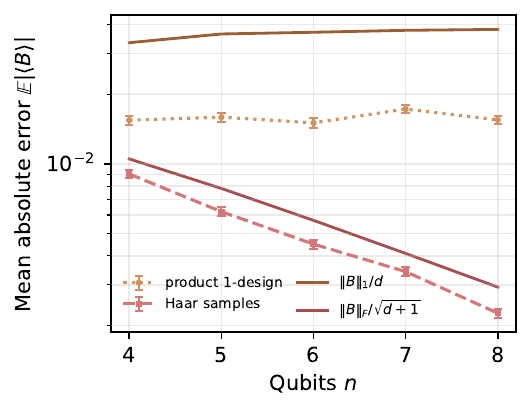}
\caption{
\textbf{The 1-design--2-design gap.}  Mean absolute error for one shallow
Trotter step.  Markers average 320 product-Pauli or Haar states; bars denote one
standard error.  Solid curves are the bounds $\norm{B(M)}_1/d$ and
$\norm{B(M)}_F/\sqrt{d+1}$ from Proposition~\ref{prop:design-simulation-gap},
evaluated for the same $B(M)$.  
}
\label{fig:observable_design_mean_abs}
\end{figure}

Our first result is the observation of an exponential gap between the observable error of 1-design and 2-design ensembles.
\begin{proposition}[Design-dependent error bounds]
\label{prop:design-simulation-gap}
For every Hermitian $M$, a pure-state 1-design obeys
$\E_{\cE}|X_\psi(M)|\le\norm{B(M)}_1/d$, while a pure-state 2-design obeys
$\E_{\cE}|X_\psi(M)|\le\norm{B(M)}_F/\sqrt{d+1}$.
\end{proposition}

A 1-design inherently satisfies $\E_{\cE}X_\psi(M)=0$, and the 1-design upper bound in Proposition~\ref{prop:design-simulation-gap} is saturated by
a uniform eigenbasis ensemble.
Thus, when the error observable satisfies $\norm{B(M)}_1=\mathcal O(d\norm{B(M)}_F)$, 1-designs permit a mean
absolute error $\mathcal O(\norm{B(M)}_F)$, whereas 2-designs suppress this error to
$\mathcal O(\norm{B(M)}_F/\sqrt d)$. 

This dimension-dependent decay differs from standard operator-norm guarantees. For example, existing product-formula bounds such as those in Ref.~\cite{li2024principal} give
$\E|X_\psi(M)|\le(t/d)\norm{[K,M]}_1$, where
$e^{-itK}=U_0U^\dagger$.  Since $\norm{[K,M]}_1$ can scale as $\mathcal O(d)$, this
bound need not decay with dimension.

Figure~\ref{fig:observable_design_mean_abs} illustrates the design-dependent
error bounds of Proposition~\ref{prop:design-simulation-gap} using Haar-random
states and the $1$-design product-Pauli ensemble $\cE_{\rm PP}$. We define $\cE_{\rm PP}$ by
sampling every qubit independently and uniformly from
$\ket0,\ket1,\ket+,\ket-,\ket{+i},\ket{-i}$.  A weight-$w$ Pauli string obeys
$\E_{\cE_{\rm PP}}|\langle P\rangle_\psi|=3^{-w}=\mathcal O(1)$ at fixed $w$, whereas
the 2-design bound in Proposition~\ref{prop:design-simulation-gap} gives
$1/\sqrt{d+1}$.
Thus $\cE_{\rm PP}$ is a concrete observation, not an assumption of the result.

Figure~\ref{fig:observable_design_mean_abs} gives a direct shallow-circuit
check of Proposition~\ref{prop:design-simulation-gap}: at every size, the
product-Pauli and Haar estimates lie below $\norm{B(M)}_1/d$ and
$\norm{B(M)}_F/\sqrt{d+1}$, respectively.  The Haar signal follows its
dimension-decaying bound, whereas the product-Pauli signal remains visible.  A
single first-order Trotter--Suzuki step
\cite{suzuki1991general} is used, so the gap already appears in a single step.
The Pauli-weight decomposition derived in the
Supplemental Material, Sec.~\ref{app:design-gap}, identifies the
low-weight channels behind the product-Pauli curve.

\textit{Frame potential.---}
To quantify the ensemble structure behind the stronger mean-absolute error
bound, we study the second moment through the \emph{second frame potential},
\begin{equation}
F_2(\cE)=\E_{\psi,\phi\sim\cE}|\langle\psi|\phi\rangle|^4
=\Tr[M_2(\cE)^2].
\end{equation}
Every pure-state 1-design obeys the exact range 
\begin{equation}
\frac{2}{d(d+1)}\le F_2(\cE)\le\frac{1}{d}.
\label{eq:frame_range}
\end{equation}
The lower endpoint is attained exactly by state 2-designs, and the upper endpoint
is attained exactly by a uniform orthonormal basis after states differing only
by global phase are combined.  Thus, among pure-state 1-designs, the largest
second frame potential exceeds the smallest by a factor $(d+1)/2$,
even though all these 1-design ensembles have the same first moment. For the product-Pauli ensemble, one can verify that
$F_2(\cE_{\rm PP})=3^{-n}$ and $F_2(\Haar)=2/[2^n(2^n+1)]$.  Section~\ref{app:frame-range} of the
Supplemental Material proves the range Eq.~\eqref{eq:frame_range} and its equality.

\begin{figure}[b]
\includegraphics[width=1\columnwidth]{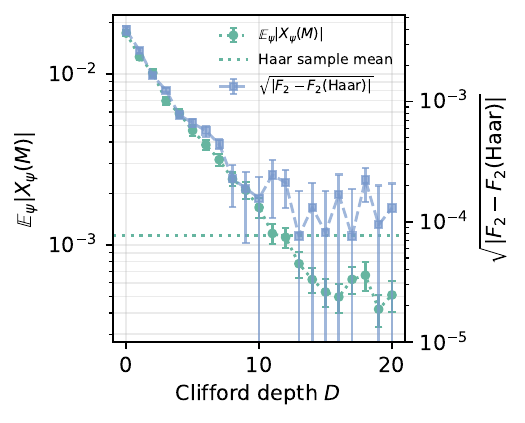}
\caption{
\textbf{Excess frame potential versus mean observable error during the random-circuit mixing.} A 10-qubit brickwork circuit of random
nearest-neighbor two-qubit Clifford gates is applied to product-Pauli inputs.
The left axis shows the mean absolute observable error, with a Haar reference;
the right axis shows
$\sqrt{|F_2-2/[d(d+1)]|}$.  Increasing depth drives the ensemble toward the
2-design scale and suppresses the selected error, but the rates need not
coincide.  Sampling details are given in the Section~\ref{app:numerics} of Supplementary Materials.
}
\label{fig:random_clifford_frame_error}
\end{figure}

\textit{Excess frame potential and residual covariance.---}
While the frame potential $F_2$
  quantifies the total magnitude of the ensemble's departure from a 2-design, it leaves the directional structure of this residual unspecified. To resolve this, we evaluate the \emph{excess frame potential}, 
  \begin{equation}
      F_2(\cE)-F_2(\Haar),
  \end{equation} by defining  the second-moment
residual $\Delta_2$ and its Pauli-basis representation $\Gamma$ by $\Delta_2:=M_2(\cE)-\frac{\I+\Swap}{d(d+1)}$ and 
\begin{equation}
\begin{split}
\Gamma_{\mu\nu}&:=\Tr[(P_\mu\otimes P_\nu)\Delta_2].
\end{split}
\end{equation}
For every pure-state ensemble,
$F_2(\cE)-F_2(\Haar)=\Tr(\Delta_2^2)$. For an observable $O$, the deviation
of its mean square from the Haar value is $\Tr[(O\otimes O)\Delta_2]$. If $\cE$ is a 1-design, $\Gamma$ is
the real symmetric Haar-subtracted covariance of Pauli expectation values and
its identity row and column vanish as $\Delta _2$ is traceless.  We keep the full $d^2\times d^2$ matrix
and use the coefficient-space normalization
$\norm{\Gamma}_F^2=d^{-2}\Tr(\Gamma^{\mathsf T}\Gamma)$.  For a nonzero
$D$-dimensional coefficient-space matrix $X$, define its normalized Frobenius norm by
$\norm{X}_F:=\sqrt{\Tr(X^{\mathsf T}X)/D}$ and its stable rank
$r_{\rm eff}(X):=D\norm{X}_F^2/\norm{X}^2 = \Tr(X^{\mathsf T}X)/\norm{X}^2$.

\begin{proposition}[Frame decomposition]
\label{thm:frame-observable-decomposition}
Let $\cE$ be a state 1-design, and $O=\sum_{\mu\ne0}o_\mu P_\mu$ a traceless Hermitian operator. The following identities hold:
$
F_2(\cE)-F_2(\Haar)=\norm{\Gamma}_F^2$ and $
\E_{\cE}|\langle O\rangle_\psi|^2
=\norm{O}_F^2/(d+1)+o^{\mathsf T}\Gamma o.
$
Moreover,
\begin{equation}
\max_{\substack{\norm{O}_F=1}}
\left|
\E_{\cE}|\langle O\rangle_\psi|^2-
\E_{\Haar}|\langle O\rangle_\psi|^2
\right|=\norm{\Gamma}.
\label{eq:opt_obs}
\end{equation}
When $\Gamma\ne0$, the stable-rank definition also gives
\begin{equation}
\norm{\Gamma}
=d\sqrt{\frac{F_2(\cE)-F_2(\Haar)}{r_{\rm eff}(\Gamma)}}.
\label{eq:frame_stable_rank}
\end{equation}
\end{proposition}

Proposition~\ref{thm:frame-observable-decomposition}
separates amount from direction.  The excess frame potential determines the total
residual weight $\norm{\Gamma}_F^2$. For a specific measurement, the resulting error is governed by the quadratic form $o^{\mathsf T}\Gamma o$, while the spectral norm $\norm{\Gamma}$ bounds the worst-case deviation along the most sensitive direction.  A small
stable rank concentrates the same total weight into fewer, more detectable
directions.  The structural identities and proof are given in the
Supplemental Material, Sec.~\ref{app:frame-decomposition}.
We also specialize this analysis to error observables reachable
by a fixed simulation and accessible measurements, giving the maximal absolute
deviation from the Haar mean-square error and the corresponding optimal
measurement (see Corollary~\ref{cor:simulation-reachable-signal} in End Matter).

Figure~\ref{fig:random_clifford_frame_error} illustrates how the excess frame potential and the error for one observable change with circuit depth.
The depth-zero ensemble is
$\cE_{\rm PP}$, and applying any
unitary to a 1-design preserves its first moment, so every depth remains a
1-design.  Sufficiently deep local random circuits converge in their moments
toward approximate designs
\cite{harrow2009,brown2010convergence,brandao2016local,schuster2025random} while in
the present Clifford setting the uniform limiting group is a unitary 3-design,
so its action generates a state 2-design
\cite{webb2015clifford,zhu2017multiqubit}.  The simultaneous
decrease of the two curves is therefore natural, while their unequal rates
reflect the distinction between total residual weight and one fixed observable
direction.

\begin{figure*}[t]
\centering
\begin{minipage}[t]{0.49\textwidth}
\centering
\textbf{(a)} Initial state: $\ket{0}^{\otimes n}$\\[-0.6ex]
\includegraphics[width=\linewidth]{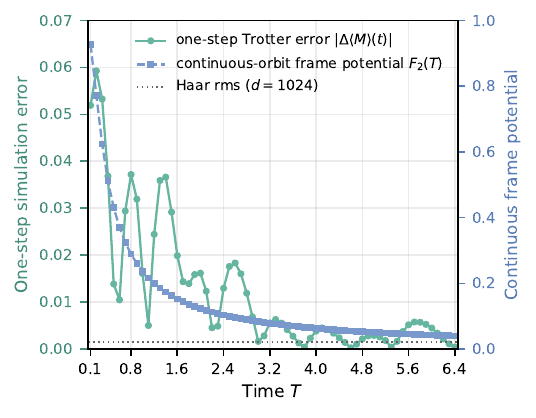}
\end{minipage}\hfill
\begin{minipage}[t]{0.49\textwidth}
\centering
\textbf{(b)} Initial state: $\ket{+}^{\otimes n}$\\[-0.6ex]
\includegraphics[width=\linewidth]{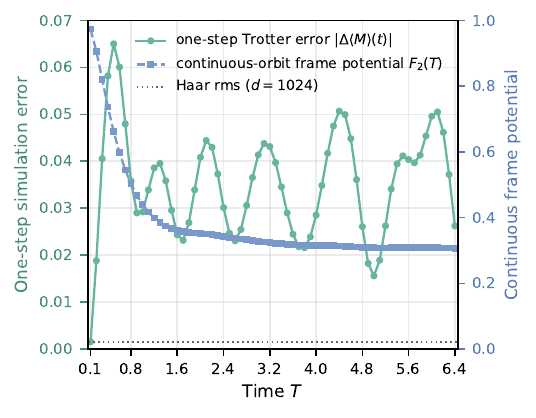}
\end{minipage}
\caption{
\textbf{Dynamical suppression of a one-step Trotter error.}  The green curve
(left axis) compares one exact and one first-order Trotter step applied to the
same exact snapshot; the blue curve (right axis) is the continuous-time frame
potential of the trajectory segment from $0$ to $T$, and the gray dotted line
is the root-mean-square (rms) error for Haar-random inputs in the full
$d=2^{10}=1024$ Hilbert space.  In (a), rapid decay of overlaps between trajectory states accompanies a $1/T$ upper bound and a strongly suppressed late one-step error.  In (b), restricted
spectral support leaves both a large frame-potential plateau and unsuppressed
directional error.  The horizontal coordinate labels exact starting snapshots,
not an accumulated-error calculation.
}
\label{fig:single-trajectory-frame-error}
\end{figure*}

\textit{Trotter errors along a single trajectory.---}
 The frame-potential method can be extended to temporal ensembles generated along a single Hamiltonian trajectory with fixed input state.
Specifically, we replace
the finite input ensemble by a continuous ensemble sampled along the evolution
trajectory~\cite{mark2024maximum}. For a fixed time-independent
$n$-qubit Hamiltonian $H$ and initial pure state $\ket{\psi_0}$, the exact
trajectory is $\ket{\psi(s)}=e^{-iHs}\ket{\psi_0}$, $0\le s\le T$.
Uniform sampling of the trajectory parameter $s$ defines the temporal ensemble
$\cE_T=\{ds/T,\ket{\psi(s)}:0\le s\le T\}$.
Replacing ensemble sums by time integrals extends the second-moment framework to $\cE_T$, without
assuming that the trajectory forms a state design.

Specializing Eq.~(9) of Ref.~\cite{hou2026state} to a single initial state
and moment order $k=2$ gives its frame potential,
\begin{align}
F_2(T):=F_2(\cE_T)
&=\frac{1}{T^2}\int_0^T\!ds\int_0^T\!du\,
|\langle\psi(s)|\psi(u)\rangle|^4
\notag\\
&=\frac{2}{T^2}\int_0^T(T-\tau)C_2(\tau)\,d\tau,
\label{eq:trajectory-frame-potential}
\end{align}
where $C_2(\tau)=|\bra{\psi_0}e^{-iH\tau}\ket{\psi_0}|^4$ is the squared
survival probability, i.e., the square of the Loschmidt echo in the
return-probability convention~\cite{campos2010unitary}.
The integral retains contributions from short-time returns.

Hou et al.~\cite{hou2026state} identify the short-time return peak
as the source of an $\mathcal O(T^{-1})$ finite-time correction.  For the fixed
trajectory considered here, we assume an integrable upper bound on returns
above their long-time average $C_2(\infty)=F_2(\infty)$.
It gives $F_2(T)\le C_2(\infty)+A_2/T$ over the time range considered, with
$T$-independent $A_2$ (see Sec.~\ref{app:continuous-trajectory} in Supplemental Material).
We use the resulting $1/T$ bound on the frame potential only while the
plateau $C_2(\infty)$ remains subleading.  Restricted spectral support can
instead sustain a large plateau.
Figure~\ref{fig:single-trajectory-frame-error} illustrates these two behaviors:
the zero-state trajectory exhibits a $1/T$ upper bound and a selected late-time
one-step PF1 error on the same $10^{-3}$ scale as the full-space Haar rms baseline,
whereas the plus-state trajectory retains both a large frame-potential plateau and
an unsuppressed observable error.  Each plotted point compares one exact
and one PF1 step from the same exact state; the figure does not measure
accumulated errors or vary the number of Trotter steps.

For a simulation of duration $T$ with $r$ Trotter steps, set $\delta t=T/r$
and $U_r(T)=[S_p(\delta t)]^r$.  The accumulated observable error from an
initial state $\ket{\psi(s)}$ is
\begin{align}
e_r(s;T)&=|\bra{\psi(s)}B_r(M;T)\ket{\psi(s)}|,    
\end{align}
where
$B_r(M;T)=U_0^\dagger(T)MU_0(T)-U_r^\dagger(T)MU_r(T)$.
We sample the initial offset only within the first step, $s\in[0,\delta t]$.
Expanding $U_0(T)-U_r(T)$ into single-step terms introduces the successive
exact states $\ket{\psi(s+j\delta t)}$ and averaging over this short offset
interval brings in the frame potential of the full trajectory.

\begin{proposition}[Trajectory-assisted Trotter scaling]
\label{thm:trajectory-trotter-scaling}
For fixed $H$ and $\ket{\psi_0}$, let $S_p$ be a $p$th-order product formula
with commutator scale $\widetilde\alpha_{\rm comm}$, and set $\delta t=T/r$.
Suppose both evolutions preserve the same invariant sector and consider a
time range in which $F_2(T)=\mathcal O(T^{-1})$, with an implicit constant independent
of $T$. Then for a fixed observable $M$, the simulation error satisfies
$
    \E_{s\sim\mathrm{Uniform}[0,\delta t]}e_r(s;T)=O\left(\dfrac{T^{p+3/4}}{r^p}\right).
$
To guarantee
$\E_{s\sim\mathrm{Uniform}[0,\delta t]}e_r(s;T)\le\epsilon$, it suffices to choose
\begin{equation}
r=O\!\left[
\left(\frac{\widetilde\alpha_{\rm comm}\norm{M}}{\epsilon}\right)^{1/p}
T^{1+3/(4p)}\right].
\label{eq:trajectory-assisted-trotter-number}
\end{equation}

\end{proposition}
The proof is given in Sec.~\ref{app:trajectory-trotter-scaling} of Supplemental Material.
The frame-potential assumption can be ensured by two physical conditions.
\emph{(i) Rapid evolution to nearly orthogonal states.}
The overlap with the initial state falls on a $T$-independent timescale
$\tau_c$ and remains small over the time range considered.
For time-independent $H$, the overlap of any two snapshots depends only on
their time separation, so the short-time return peak contributes a strip
of width $\mathcal O(\tau_c)$ around $s=u$ in Eq.~\eqref{eq:trajectory-frame-potential}.
This keeps returns small in the selected window.
\emph{(ii) An appropriate time range.}
Take $T$ beyond the initial overlap decay, but with $C_2(\infty)\ll A_2/T$
and before substantial recurrences.  At very short times the snapshots
still strongly overlap, so $F_2(T)\simeq1$; at long times a finite-size
plateau or renewed returns can end the $1/T$ regime.
Requiring $C_2(\infty)+A_2/T<1$ also makes the estimate nontrivial.
These conditions explain the intermediate time range in which the proposition's
assumption is useful; they do not require exact or permanent orthogonality.
Section~\ref{app:continuous-trajectory} of the Supplemental Material gives the quantitative time limits. 

Without the temporal suppression, the conventional state-independent estimate
scales as $r=O[(\widetilde\alpha_{\rm comm}\norm{M}/\epsilon)^{1/p}T^{1+1/p}]$
\cite{childs2021theory}; PF1 thus changes from $T^2$ to $T^{7/4}$ in the stated
window.  Here $\epsilon$ is a tolerance for the mean absolute accumulated
error in expectation values, averaged only over $s\in[0,\delta t]$.
The proposition bounds this statistical quantity through an rms upper bound;
it is not a worst-case error guarantee or a finite-sample uncertainty.
Proposition~\ref{prop:accumulated-trajectory-bound} in Sec.~\ref{app:trajectory-trotter-scaling} of
Supplemental Material gives
the quantitative bound, its single-step error and saturation conditions.
The same section extends the bound to each fixed initial state using the
discrete trajectory frame potential and its comparison with $F_2(T)$
[Eqs.~\eqref{eq:discrete-trajectory-frame} and~\eqref{eq:trajectory-frame-sampling}].
Numerical PF2 step counts versus evolution time for fixed initial states are presented in the
End Matter (see Fig.~\ref{fig:single-trajectory-pf2-resources}).

\textit{Discussion and Conclusion.---}
Our work reveals how much randomness may hide quantum simulation (Trotter) error in Hamiltonian simulation.  A 2-design yields a dimension-suppressed upper bound on the mean
absolute observable error, with the exponential ($\sqrt d$) improvement over the 1-design
bound under the norm scaling in Proposition~\ref{prop:design-simulation-gap}.
This suppression reduces the number of Trotter steps sufficient to reach a
target accuracy, but makes an
implementation error harder to diagnose. Therefore, one may not use 2-designs to benchmark quantum simulation error since the average error is exponentially small.

The established temporal frame-potential analysis~\cite{hou2026state}
supplies a dynamical input to our accumulated-error bound.
Proposition~\ref{thm:trajectory-trotter-scaling} uses successive exact states
in a telescoping bound on the accumulated error, with initial offsets
averaged only over $[0,\delta t]$.  The sufficient scaling
$r=\mathcal O(T^{1+3/(4p)})$ applies within the $1/T$ regime, with a resource advantage
only when the single-step error spectrum permits a nontrivial bound.  The derived step-count scaling is a sufficient upper bound rather than a tight prediction of the required resources, consistent with the slower growth observed for fixed inputs in Fig.~\ref{fig:single-trajectory-pf2-resources}.
Its rms
baseline and the finite-size frame-potential plateau terminate this improvement.
Restricted spectral support provides a further obstruction, illustrated by
the plus-state trajectory.  Thus the frame potential tells
how much second-moment structure remains, the residual spectrum tells where it
lies, and the Heisenberg Trotter-error map tells what an observable benchmark
can reach.  None of these state-dependent reductions changes the worst-case
operator-norm error of the Trotterized unitary.

Our approaches and analysis may also be extended to perturbation or coherent errors in analog quantum simulation~\cite{RevModPhys.86.153,Fengdoi:10.1126/sciadv.aef3416,PhysRevLett.119.240502}.
For a fixed Hermitian Hamiltonian mismatch $\delta H$, take
$U_0=e^{-iHt}$ and $U=e^{-i(H+\delta H)t}$. 
This perspective complements studies of
observable-dependent error sensitivity~\cite{poggi2020quantifying} and
entanglement-dependent resilience against coherent Hamiltonian
perturbations~\cite{Fengdoi:10.1126/sciadv.aef3416}.

\begin{acknowledgments}
The authors thank the support of the CCF-Tencent Rhino-Bird Open Research Fund (No. T105-TEG-2024080200001).  Q.Z. acknowledges funding from Quantum Science and Technology-National Science and Technology Major Project 2024ZD0301900, National Natural Science Foundation of China (NSFC) via Project No. 12347104 and No. 12305030, Hong Kong Research Grant Council (RGC) via No.17310926, N\_HKU718/23, and R6010-23, the Guangdong Provincial Quantum Science Strategic Initiative (Grants No.GDZX2503001).  ChatGPT and Gemini were used during the preparation of this manuscript to mainly assist with grammar, wording, coding, and sentence structure. All conceptual content, arguments, and results were developed by
the authors, with helpful discussion with AI. AI-assisted edits were reviewed and verified by the authors, who take full responsibility for the manuscript.
\end{acknowledgments}

\bibliography{ref}
\clearpage
\section*{End Matter}

\textit{Reachable upper bound of observable error.---}
The maximization in Eq.~\eqref{eq:opt_obs} treats $O$ as freely variable over
the unit sphere of traceless Hermitian operators.  Its
optimizer therefore identifies the strongest direction in the ensemble's
second-moment anisotropy, but that direction need not be an error observable
that a fixed simulation can produce.  Once $U_0$ and $U$ are fixed, the
experimenter chooses $M$ from an \emph{accessible measurement space} $\mathsf M$,
whereas the ensemble sees only the induced error observable
$B(M)=\cL(M)$.  The reachable error observables are consequently restricted to
$\cL(\mathsf M)$, generally a proper subspace of the traceless operator space.
The normalization in Eq.~\eqref{eq:opt_obs} is imposed on $O$, whereas
here we normalize the measured observable $M$.  Even when $\norm{M}_F=1$,
$\norm{B(M)}_F$ depends on $M$, and some nonzero observables satisfy
$B(M)=0$.  Thus the excess frame potential fixes
only the total residual weight, while $\norm{\Gamma}$ identifies its strongest
unconstrained direction; neither guarantees that the particular simulation
couples an accessible measurement to that direction.  The operational question
is instead how much of $\Gamma$ survives after both restrictions.

If $m$ and $b$ are the Pauli coefficient vectors of $M$ and $B(M)$, define the
real matrix $L_{\alpha\mu}:=d^{-1}\Tr[P_\alpha\cL(P_\mu)]$, so $b=Lm$.
Let $\Pi_{\mathsf M}$ be the orthogonal coefficient projector onto $\mathsf M$.
On the image space of $\Pi_{\mathsf M}$, define the real symmetric matrix
\begin{align}
A_{\rm sim}&:=\Pi_{\mathsf M}L^{\mathsf T}\Gamma L\Pi_{\mathsf M}.    
\end{align}
For an accessible observable $M$, its coefficient vector satisfies
$m=\Pi_{\mathsf M}m$, and $m^{\mathsf T}A_{\rm sim}m$ equals the difference
between its mean-square simulation error and the Haar value.
We define
\begin{equation}
D_{\rm sim}:=
\max_{\substack{M\in\mathsf M\\ \norm{M}_F=1}}
\left|\mathcal V_{\cE}(M)-\mathcal V_{\Haar}(M)\right|
\label{eq:dsim-def}
\end{equation}
where, recall, that $\mathcal V_{\cE}(M):=\E_{\psi\sim\cE}\abs{X_\psi(M)}^2$.
Here $D_{\rm sim}$ measures the largest absolute deviation from the
Haar mean-square error over normalized accessible observables.
Thus, $D_{\rm sim}=0$ means that these mean-square errors equal their Haar
values; it does not imply that the simulation error itself vanishes.

\begin{corollary}[Reachable signal]
\label{cor:simulation-reachable-signal}
Let $\cE$ be a pure-state ensemble, and $\mathsf M$ a space of accessible measurements.  For every Hermitian $M\in \mathsf M$,
\begin{align}\label{eq:nu_e}
 \mathcal V_{\cE}(M)&=m^{\mathsf T}L^{\mathsf T}Lm/(d+1)
+m^{\mathsf T}L^{\mathsf T}\Gamma Lm   
\end{align}
 and
$D_{\rm sim}=\norm{A_{\rm sim}}$.
If $A_{\rm sim}\ne0$, a normalized eigenvector $m_\ast$ associated with its
largest-magnitude eigenvalue defines an optimal observable
$M_\ast=\sum_\mu(m_\ast)_\mu P_\mu$.
\end{corollary}
The proof is given in the Supplemental Material, Section~\ref{app:reachable-signal}.
Geometrically, $A_{\rm sim}$ pulls the ensemble anisotropy back to the
accessible measurement space.  Its dominant eigenvector therefore specifies a
measurement that can actually be selected, rather than merely an abstract
witness $O$.

By applying our stable-rank identity to this filtered residual, we can bound the maximum possible deviation $D_{\rm sim}$. For any $\Gamma\ne0$, the worst-case error is constrained by:
\begin{equation}
D_{\rm sim}
\le \norm{L\Pi_{\mathsf M}}^2d
\sqrt{\frac{F_2(\cE)-F_2(\Haar)}{r_{\rm eff}(\Gamma)}}.
\label{eq:frame_bound_sim}
\end{equation}

\begin{figure}[t]
\includegraphics[width=\columnwidth]{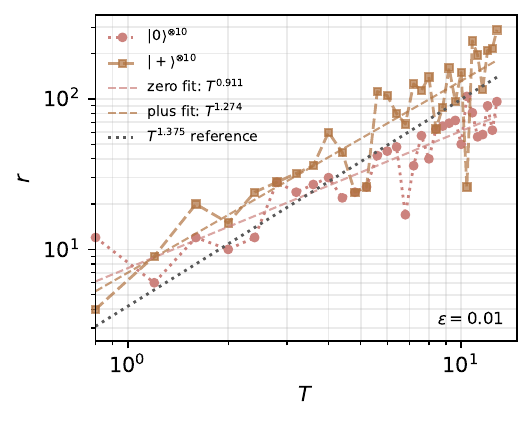}
\caption{PF2 steps for fixed initial states $\ket{0}^{\otimes10}$ and
$\ket{+}^{\otimes10}$ at absolute observable-error tolerance $\epsilon=0.01$,
with the fixed Hamiltonian and observable of
Fig.~\ref{fig:single-trajectory-frame-error} and initial offset $s=0$.
At each $T$, $r$ is the smallest scanned step count for which
the error stays below $\epsilon$ at every larger scanned count.
Solid curves are log--log least-squares fits $r=CT^\beta$ to all 31 points
per state over $0.8\le T\le12.8$.
The gray dotted $T^{1.375}$ line is a fixed-normalization reference,
not a fitted curve or an error bound.}
\label{fig:single-trajectory-pf2-resources}
\end{figure}

The decomposition of $\mathcal{V}_\cE$ in \eqref{eq:nu_e} holds for any pure-state ensemble without requiring a 1-design assumption.  Since $B(M)$ is traceless,
its identity Pauli coefficient vanishes ($b_0=d^{-1}\Tr B(M)=0$), so every term in
$b^{\mathsf T}\Gamma b$ with an identity index vanishes.  Corollary~\ref{cor:simulation-reachable-signal} splits the mean-square simulation error $\mathcal V_{\cE}(M)$ into two distinct contributions. The first term represents the fundamental, Haar-suppressed baseline. The second term isolates the state-dependent residual. Notice that this residual is shaped by two sequential filters: the physical error map ($L$) and the restricted measurement subspace ($\Pi_{\mathsf M}$).  

When $\Gamma=0$, the mean-square errors trivially match their Haar values, giving $D_{\rm sim}=0$. Alternatively, if
$q_{\mathsf M}=\operatorname{rank}(\Pi_{\mathsf M})$ and
$A_{\rm sim}\ne0$, then
$D_{\rm sim}=\sqrt{q_{\mathsf M}}\norm{A_{\rm sim}}_F/
\sqrt{r_{\rm eff}(A_{\rm sim})}$.  The magnitude of these bounds on $D_{\rm sim}$ depends on the structure of the error map $\cL(M)$, which can vanish for certain measurements. These `blind' directions -- observables $M$ for which the error map $\cL(M)$ vanishes -- are ensemble
independent: with $V=UU_0^\dagger$,
$\cL(M)=U_0^\dagger(M-V^\dagger M V)U_0$, and hence
$\cL(M)=0\Longleftrightarrow[M,V]=0$.  For such an observable, the ideal
and implemented expectation values agree for every input state.

\textit{{Numerical tests for Trotter-step counts.---}}
Figure~\ref{fig:single-trajectory-pf2-resources} compares the accumulated-error
step counts for the two fixed initial states using symmetric PF2 at
$\epsilon=0.01$, with $H$ and $M$ unchanged and no initial-offset average.
The zero state generally requires fewer steps.  Descriptive power-law fits
over $0.8\le T\le12.8$ give exponents $0.911$ and $1.274$ for the zero and
plus states, respectively.  The $T^{11/8}$ line is a slope reference.

\clearpage
\onecolumngrid
\appendix
\setcounter{secnumdepth}{1}

\section*{Supplementary Materials}

\section{Proof of Proposition~\ref{prop:design-simulation-gap}}
\label{app:design-gap}

We begin with the state-design properties needed for the proof, then explain how they control the observable error.

\paragraph{Pure-state ensembles and $k$-designs.}
In a pure-state ensemble $\cE=\{p_a,\ket{\psi_a}\}_a$, the normalized state
$\ket{\psi_a}\in\cH$ is prepared with probability $p_a$,
where $p_a\ge0$ and $\sum_a p_a=1$.  Write
$\rho_a=\ket{\psi_a}\bra{\psi_a}$ for its density operator.   The Haar ensemble is the uniform distribution over pure states:
its distribution is unchanged by applying any fixed unitary to every state.
A state design reproduces specified averages of this uniform distribution
using a possibly discrete ensemble~\cite{ambainis2007quantum,scott2006tight}.

For an integer $k\ge1$, the moment operator is $M_k(\cE):=\E_{\psi\sim\cE}\rho_\psi^{\otimes k}$.
Here $\rho_\psi^{\otimes k}$ represents the tensor product of $k$ copies of the sampled state.  An ensemble is an exact \emph{state $k$-design} when it reproduce the first $k$ moment operator of a Haar ensemble:
\begin{equation}
M_k(\cE)=M_k(\Haar)
=\frac{\Pi_{\rm sym}^{(k)}}{\binom{d+k-1}{k}}.
\label{eq:supp-state-k-design}
\end{equation}
The operator $\Pi_{\rm sym}^{(k)}$ projects onto the subspace of
$\cH^{\otimes k}$ unchanged by permutations of the copies.  By definition, a $(k+1)$-design ensemble is also $k$-design.  All vectors
$\ket{\psi}^{\otimes k}$ belong to this symmetric subspace; the denominator
is its dimension and gives the Haar moment unit trace~\cite{scott2006tight}.
Specially, for a 1-design ensemble $\cE_1$, $M_1(\cE_1)=\I/d$, and for a 2-design ensemble $\cE_2$, $M_2(\cE_2)=\dfrac{\I+\Swap}{d(d+1)}$, where $\I$ represents the identity operator and $\Swap$ represents the swap operator on $\cH\otimes\cH$.  In our case, we are interested in the first two moment of the error operator $B(M)=U_0^\dagger MU_0-U^\dagger MU$, which satisfies $\Tr[B(M)]=0$ and is Hermitian when $M$ is Hermitian.
For any input ensemble,
$\E_{\cE}X_\psi(M)=\Tr[B(M)M_1(\cE)]$.
To express the mean square using two copies, we use the tensor-product trace
identity $\Tr(A)\Tr(B)=\Tr(A\otimes B)$, valid for any square matrices $A,B$.
With $\rho_\psi=\ket{\psi}\bra{\psi}$, applying this identity to
$A=B=B(M)\rho_\psi$ gives
$\mathcal V_{\cE}(M)=\E_\cE[\Tr(B(M)\rho_\psi)\Tr(B(M)\rho_\psi)]=\Tr[(B(M)\otimes B(M))M_2(\cE)]$.

\paragraph{1-design bound.} Since $B(M)$ is
Hermitian, $|\langle B(M)\rangle_\psi|\le\langle|B(M)|\rangle_\psi$. Averaging over a 1-design gives
$\E_{\cE} \langle |B(M)| \rangle_\psi = \Tr(|B(M)|)/d = \norm{B(M)}_1/d$,
proving the 1-design bound in
Proposition~\ref{prop:design-simulation-gap}~\cite{li2024principal}.  Every vector
in an eigenbasis of $B(M)$ saturates this inequality, and the uniform
eigenbasis ensemble is a 1-design, proving tightness. 
\paragraph{2-design bound.}
The swap operator obeys $\Tr[\Swap(A\otimes B)]=\Tr(AB)$ for any operators
$A,B$ on $\cH$.  Together with cyclicity of the trace, this identity
evaluates the contribution containing $\Swap$.
For every state 2-design and Hermitian $M$,
\begin{equation}
\begin{aligned}
\mathcal V_{\cE}(M)&=\Tr\left[(B(M)\otimes B(M))\dfrac{\I+\Swap}{d(d+1)}\right]\\
&=\Tr\left[\dfrac{B(M)\otimes B(M)}{d(d+1)}\right]+\Tr[\dfrac{B(M)^2}{d(d+1)}]\\&=
\frac{\norm{B(M)}_F^2}{d+1}.
\label{eq:supp-design-haar-error}
\end{aligned}
\end{equation}
The last equality uses
$\Tr[B(M)\otimes B(M)]=(\Tr[B(M)])^2=0$ and
$\norm{B(M)}_F^2=\Tr[B(M)^2]/d$.
Applying Jensen's inequality ($\E|X| \le \sqrt{\E[X^2]}$) then proves the mean-absolute bound in
Proposition~\ref{prop:design-simulation-gap}.
Combining tightness with the stated trace-to-Frobenius
norm scaling proves the final separation.

\paragraph{Product-Pauli ensemble.} Let $r_\alpha=\langle\sigma_\alpha\rangle_\psi$ be a one-qubit Bloch
component.  Uniform sampling over the six single-qubit Pauli eigenstates gives
$\E [r_\alpha]=0$ and
$\E[r_\alpha r_\beta]=\delta_{\alpha\beta}/3$.  Because the qubits are sampled independently, the expectation of a multi-qubit Pauli string factorizes, yielding
$\E_{\cE_{\rm PP}}[\langle P\rangle_\psi\langle Q\rangle_\psi]
=\delta_{PQ}3^{-\operatorname{wt}(P)}$.
For a traceless observable $O=\sum_{P\ne\I}o_P P$, this gives
$\E_{\cE_{\rm PP}}\langle O\rangle_\psi^2
=\sum_{P\ne\I}3^{-\operatorname{wt}(P)}o_P^2$.  A weight-$w$ Pauli
expectation takes values $0,\pm1$, giving mean absolute value $3^{-w}$ while the
Jensen's inequality bounds its 2-design value by $1/\sqrt{d+1}$.


\section{Frame-potential range}
\label{app:frame-range}

The operator $M_2(\cE)$ is positive, has unit trace, and is supported on the
symmetric subspace of dimension $d(d+1)/2$.  Cauchy--Schwarz therefore gives
\begin{equation}
F_2(\cE)=\Tr[M_2(\cE)^2]\ge\frac{2}{d(d+1)},
\label{eq:supp-frame-lower-bound}
\end{equation}
with equality exactly when
$M_2(\cE)=2\Pi_{\rm sym}/[d(d+1)]=\frac{\I+\Swap}{d(d+1)}$, i.e., when $\cE$ is a
2-design.  For a 1-design, writing
$z=|\langle\psi|\phi\rangle|^2\in[0,1]$ gives
\begin{equation}
F_2(\cE)=\E z^2\le\E z=\Tr[M_1(\cE)^2]=\frac1d.
\label{eq:supp-frame-upper-bound}
\end{equation}
Equality holds if and only if $z\in\{0,1\}$ almost surely, so any two states
of nonzero probability are either orthogonal or identical up to a global phase.
For each group of states differing only by a global phase, retain one
representative $\ket{\phi_j}$ and assign it the summed probability $p_j$.
This leaves all ensemble moments unchanged.  The representatives are mutually
orthogonal, and $M_1=\sum_j p_j\ket{\phi_j}\bra{\phi_j}=\I/d$ forces exactly $d$ representatives, each with $p_j=1/d$.
This proves Eq.~\eqref{eq:frame_range} and both equality conditions.

\section{Proof of Proposition~\ref{thm:frame-observable-decomposition}}
\label{app:frame-decomposition}

For any pure-state ensemble, $M_2(\cE)$ and $\frac{\I+\Swap}{d(d+1)}$ have unit trace and
are supported on the symmetric subspace.  Because $\frac{\I+\Swap}{d(d+1)}$ is proportional
to the identity there, $\Tr(\frac{\I+\Swap}{d(d+1)}\Delta_2)=0$.  Pauli completeness gives
$\Delta_2=d^{-2}\sum_{\mu,\nu}\Gamma_{\mu\nu}P_\mu\otimes P_\nu$.
The normalization $d^{-1}\Tr(P_\mu P_\alpha)=\delta_{\mu\alpha}$
implies
$\Tr[(P_\mu\otimes P_\nu)(P_\alpha\otimes P_\beta)]
=d^2\delta_{\mu\alpha}\delta_{\nu\beta}$.
Thus, squaring the expansion and taking the trace removes all unequal
index pairs and gives
$\Tr(\Delta_2^2)=d^{-2}\sum_{\mu,\nu}\Gamma_{\mu\nu}^2$.
Since $\Gamma$ is real, the sum is $\Tr(\Gamma^{\mathsf T}\Gamma)$, yielding
\begin{equation}
F_2(\cE)-F_2(\Haar)=\Tr(\Delta_2^2)
=d^{-2}\Tr(\Gamma^{\mathsf T}\Gamma)=\norm{\Gamma}_F^2.
\label{eq:supp-frame-excess-moment}
\end{equation}

The matrix $\Gamma$ is real because it pairs Hermitian operators and symmetric
because $\Delta_2$ is invariant under exchange of its two copies.  For a
1-design,
$\Tr_1\Delta_2=\Tr_2\Delta_2=0$, which gives
$\Gamma_{0\nu}=\Gamma_{\mu0}=0$.  On the subspace of traceless operators,
\begin{equation}
\Gamma_{\mu\nu}
=\E_{\cE}[\langle P_\mu\rangle_\psi\langle P_\nu\rangle_\psi]
-\frac{\delta_{\mu\nu}}{d+1}.
\label{eq:supp-pauli-covariance-residual}
\end{equation}
Purity also implies
$\sum_{\mu\ne0}\langle P_\mu\rangle_\psi^2=d-1$; subtracting the Haar value
and averaging yields $\Tr\Gamma=0$.  Thus every nonzero $\Gamma$ has both
positive and negative eigenvalues.

For $O=\sum_{\mu\ne0}o_\mu P_\mu$, substitution gives
\begin{equation}
\E_{\cE}|\langle O\rangle_\psi|^2
=\Tr[(O\otimes O)(\frac{\I+\Swap}{d(d+1)}+\Delta_2)]
=\frac{\norm{O}_F^2}{d+1}+o^{\mathsf T}\Gamma o.
\label{eq:supp-observable-frame-decomposition}
\end{equation}
Since $\norm{O}_F^2=\norm{o}_2^2$ and the identity block of $\Gamma$ vanishes,
Rayleigh--Ritz gives Eq.~\eqref{eq:opt_obs}.  Finally,
Eq.~\eqref{eq:frame_stable_rank} combines the definition of $r_{\rm eff}$ with
the frame-potential identity in
Proposition~\ref{thm:frame-observable-decomposition}.

\section{Proof of Corollary~\ref{cor:simulation-reachable-signal}}
\label{app:reachable-signal}

Directly from $M_2=\frac{\I+\Swap}{d(d+1)}+\Delta_2$, every pure-state ensemble obeys
\begin{equation}
\mathcal V_{\cE}(M)=\frac{\norm{B(M)}_F^2}{d+1}+b^{\mathsf T}\Gamma b.
\label{eq:supp-reachable-error-decomposition}
\end{equation}
Because $B(M)$ is traceless and $b=Lm$, terms involving the identity operator do not contribute;
using $\norm{B(M)}_F^2=m^{\mathsf T}L^{\mathsf T}Lm$ proves the expression for
$\mathcal V_{\cE}(M)$ in Corollary~\ref{cor:simulation-reachable-signal}.  For
$M\in\mathsf M$,
$m=\Pi_{\mathsf M}m$,
so its residual deviation is $m^{\mathsf T}A_{\rm sim}m$.  The restricted matrix
$A_{\rm sim}$ is real symmetric; Rayleigh--Ritz therefore proves
$D_{\rm sim}=\norm{A_{\rm sim}}$ and the optimizer statement.  An identity
component is a harmless null direction because $\cL(\I)=0$.

The bound in Eq.~\eqref{eq:frame_bound_sim} follows from
submultiplicativity, the identity
$F_2(\cE)-F_2(\Haar)=\norm{\Gamma}_F^2$, and the definition of $r_{\rm eff}$.
For the condition of zero observable error, $V=UU_0^\dagger$ gives
$\cL(M)=U_0^\dagger(M-V^\dagger M V)U_0$.
Consequently $\cL(M)=0$ iff $M=V^\dagger M V$, equivalently iff $[M,V]=0$.

\section{Temporal frame potential and the applicable time window}
\label{app:continuous-trajectory}

Equation~\eqref{eq:trajectory-frame-potential} is the single-state, $k=2$
case of Eq.~(9) in Ref.~\cite{hou2026state}.  That work analyzes the
short-time origin of $1/T$ finite-time corrections and derives their scaling
for GUE-averaged frame potentials in its Supplemental Material, Appendix~5.
We define the long-time average of $C_2$ by
\begin{equation}
C_2(\infty):=\lim_{T'\to\infty}\frac{1}{T'}\int_0^{T'}C_2(\tau)\,d\tau.
\label{eq:trajectory-return-plateau}
\end{equation}
As derived from the temporal ensemble limit, this identically matches the asymptotic frame potential, $F_2(\infty) = C_2(\infty)$.
For our fixed Hamiltonian, a sufficient condition is the upper bound
$C_2(\tau)\le C_2(\infty)+g(\tau)$ for $0\le\tau\le T$, with $g\ge0$ and
$2\int_0^T g(\tau)d\tau\le A_2$, where the bounding constant $A_2$ is independent of $T$ across the selected time window.  
The cited integral identity immediately gives
\begin{equation}
F_2(T)\le C_2(\infty)+\frac{A_2}{T}.    
\label{eq:supp-temporal-frame-upper-bound}
\end{equation}

Proposition~\ref{thm:trajectory-trotter-scaling} requires the total frame potential to scale as $F_2(T)=\mathcal O(T^{-1})$. Therefore, the upper bound in Eq.~\eqref{eq:supp-temporal-frame-upper-bound} satisfies this prerequisite only while the permanent plateau remains subleading ($C_2(\infty)\lesssim A_2/T$).
A $1/T$ correction to a substantial plateau
alone does not suffice.  For a small plateau $0<C_2(\infty)\ll1$, a
nontrivial bound dominated by the $1/T$ term is obtained after the initial
overlap decay, specifically in the window
\begin{equation}
    A_2/[1-C_2(\infty)]<T\ll A_2/C_2(\infty).
\end{equation}
The lower limit ensures $F_2(T)\le C_2(\infty)+A_2/T<1$; the upper limit keeps the
plateau much smaller than $A_2/T$.
Finite dimension prevents this regime from extending to arbitrarily large
$T$, since $F_2(T)\ge2/[d(d+1)]$.


To determine whether a specific initial state exhibits a sufficiently small plateau to access this $1/T$ regime, we must evaluate $C_2(\infty)$ in the energy eigenbasis. We assume second-order nonresonance among the occupied energy levels
: $E_\lambda+E_\mu=E_\nu+E_\xi$ holds only when the two
energy pairs coincide up to permutation.  Under this condition, for the single initial state $\ket{\Psi_0}$, Ref.~\cite{hou2026state} gives
\begin{equation}
    F_2(\infty)=2\sum_{n_1,n_2}\dfrac{|\langle\psi_{n_1}|{\Psi_0}\rangle|^4|\langle\psi_{n_2}|{\Psi_0}\rangle|^4}{\Omega({n_1,n_2})},
\end{equation}
where $\Omega(n_1,n_2)=1+\delta_{n_1,n_2}$ and $\ket{\psi_{n_1}},\ket{\psi_{n_2}}$ denote the energy eigenstates. 
Define the spectral weights
$P_\lambda=|\langle\psi_\lambda|\Psi_0\rangle|^2$, we obtain the plateau

\begin{equation}
\begin{aligned}
C_2(\infty)&=2\sum_{\lambda_1\ne \lambda_2}P_{\lambda_1}^2P_{\lambda_2}^2+\sum_{\lambda}P_\lambda^4\\
&=2\left(\sum_\lambda P_\lambda^2\right)^2
-\sum_\lambda P_\lambda^4.
\label{eq:supp-spectral-return-plateau}
\end{aligned}
\end{equation}
This matches the infinite-time average of the squared Loschmidt echo 
computed in Ref.~\cite{campos2010unitary}. Define the effective dimension of the initial state in the energy basis by $D_{\text{eff}}=1/(\sum_\lambda P_\lambda^2)$. Since
$0<\sum_\lambda P_\lambda^4\le(\sum_\lambda P_\lambda^2)^2$,
Eq.~\eqref{eq:supp-spectral-return-plateau} gives
\begin{equation}
    \dfrac1{D_{\text{eff}}^2}\le C_2(\infty)<\dfrac{2}{D_{\text{eff}}^2}.
    \label{eq:plateau-vs-effective-dimension}
\end{equation}

Equation~\eqref{eq:plateau-vs-effective-dimension} demonstrates that the plateau size is governed by how broadly the initial state is distributed across the energy spectrum. If the state is spread across a large number of energy levels, the individual weights $P_\lambda$ are small, making $D_{\text{eff}}$ dimensionally large and yielding a vanishingly small plateau. Conversely, an initial state with restricted spectral support is concentrated in a few energy levels, resulting in a large $C_2(\infty)$ plateau that blocks the $1/T$ suppression regime. This mathematically underpins the divergent behaviors of the zero-state and plus-state trajectories shown in Fig.~\ref{fig:single-trajectory-frame-error}.

\section{Accumulated-error bound and proof of Proposition~\ref{thm:trajectory-trotter-scaling}}
\label{app:trajectory-trotter-scaling}

Set $\delta t=T/r$ and use the main-text notation
$U_0(t)=e^{-iHt}$ and $U_r(T)=[S_p(\delta t)]^r$ for the exact and
Trotterized evolutions.  Define the single-step difference
$D=U_0(\delta t)-S_p(\delta t)$.
We use $\norm{X}$ for the operator norm and
$\norm{X}_F^2=\Tr(X^\dagger X)/N$ for the normalized Frobenius norm
on an $N$-dimensional space.  The conversion factor to
$[\Tr(X^\dagger X)]^{1/2}$ is $\sqrt d$ on one copy and $d$ on the
full two-copy space.
For $s\sim\mathrm{Uniform}[0,\delta t]$, write
$\bar e_{\delta t}=\E_s e_r(s;T)$ and
$\epsilon_{{\rm rms},\delta t}=[\E_s e_r(s;T)^2]^{1/2}$.
The following bound is valid for every pure-state trajectory; the frame-potential
assumption of Proposition~\ref{thm:trajectory-trotter-scaling} is needed only
for its resource implication.

\begin{proposition}[Accumulated trajectory error]
\label{prop:accumulated-trajectory-bound}
For fixed $H$, $\ket{\psi_0}$, and Hermitian $M$, set 
$D=U_0(\delta t)-S_p(\delta t)$ and $Q=D^\dagger D$ and define the single-step error coefficient
\begin{equation}
\kappa_{\delta t}:=\frac12\sqrt{(d+2)\Tr(Q^2)+(\Tr Q)^2}.
\label{eq:local-defect-frame-scale}
\end{equation}
Then the accumulated error satisfies
\begin{align}
\bar e_{\delta t}&\le\epsilon_{{\rm rms},\delta t}
\notag\\
&\le2\norm{M}r\min\!\left\{\norm{D},
\sqrt{\kappa_{\delta t}}\,F_2(T)^{1/4}\right\}.
\label{eq:accumulated-trajectory-bound}
\end{align}
If $U_0(\delta t)$ and $S_p(\delta t)$ both preserve a subspace
containing the initial state, all
quantities may be evaluated within that sector, with its dimension $d$.
\end{proposition}

\textit{Proof of the averaged bound and Proposition~\ref{thm:trajectory-trotter-scaling}.---}
Write $\rho(t)=\ket{\psi(t)}\bra{\psi(t)}$ and first bound the state-vector
error $a_r(s)=\norm{[U_0(T)-U_r(T)]\ket{\psi(s)}}_2$.
Expanding the difference between the two evolutions gives the exact identity 
\begin{equation}
\begin{aligned}
[U_0(T)-U_r(T)]\ket{\psi(s)}
&=\sum_{j=0}^{r-1}\left\{[S_p(\delta t)]^{r-j-1}U_0((j+1)\delta t)-[S_p(\delta t)]^{r-j}U_0(j\delta t)\right\}\ket{\psi(s)}\\
&=\sum_{j=0}^{r-1} [S_p(\delta t)]^{r-j-1}[U_0(\delta t)-S_p(\delta t)]U_0(j\delta t)\ket{\psi(s)}\\
&=\sum_{j=0}^{r-1}[S_p(\delta t)]^{r-1-j}D\ket{\psi(s+j\delta t)}.
\label{eq:trajectory-telescoping}
\end{aligned}
\end{equation}
Unitarity, the triangle inequality, and Cauchy--Schwarz give
$a_r(s)^2\le r\sum_j\Tr[Q\rho(s+j\delta t)]$.
Averaging over $s\in[0,\delta t]$ joins the $r$ adjacent intervals into
$[0,T]$.  With $\bar\rho_T=T^{-1}\int_0^T\rho(t)\,dt$, this yields
\begin{equation}
\E_s a_r(s)^2\le\frac{r}{\delta t}\int_0^T\Tr[Q\rho(t)]\,dt
=r^2\Tr(Q\bar\rho_T).
\label{eq:accumulated-local-moment}
\end{equation}
Thus the full trajectory enters through the intermediate states of one
simulation, while its initial offset is sampled only within one step.

To bound this mean-square single-step error using the frame potential, use
the temporal second moment
$M_2(\cE_T)=T^{-1}\int_0^T\rho(t)^{\otimes2}\,dt$, together with
$P_{\rm sym}=(\I+\Swap)/2$ and
$\mathcal A_Q=\tfrac12(Q\otimes\I+\I\otimes Q)P_{\rm sym}$.
Every $\rho(t)^{\otimes2}$ is supported on the symmetric subspace, so
$\Tr[\mathcal A_Q M_2(\cE_T)]=\Tr(Q\bar\rho_T)$.
The identity $\Tr[(X\otimes Y)\Swap]=\Tr(XY)$ and $P_{\text{sym}}^2=P_{\text{sym}}$ gives
\begin{equation}
\begin{aligned}
d\norm{\mathcal A_Q}_F&=\sqrt{\Tr(\mathcal A_Q^2)}\\
&=\dfrac12\sqrt{\Tr\left[(Q^2\otimes I+I\otimes Q^2+2Q\otimes Q)\dfrac{\I+\Swap}2\right]}\\
&=\frac12\sqrt{(d+2)\Tr(Q^2)+(\Tr Q)^2}\\
&=\kappa_{\delta t}.
\label{eq:defect-symmetric-lift}
\end{aligned}
\end{equation}
The factor $d=\sqrt{d^2}$ accounts for the full two-copy dimension.
Since $d\norm{M_2(\cE_T)}_F=\sqrt{F_2(T)}$, Cauchy--Schwarz gives
$\Tr(Q\bar\rho_T)\le d^2\norm{\mathcal A_Q}_F\norm{M_2(\cE_T)}_F
=\kappa_{\delta t}\sqrt{F_2(T)}$.
Also, $\norm{D}$ is the worst-case single-step state-vector error, so
$Q\le\norm{D}^2\I$ and $\Tr(Q\bar\rho_T)\le\norm{D}^2$.
Combining these bounds with Eq.~\eqref{eq:accumulated-local-moment},
$e_r(s;T)\le2\norm{M}a_r(s)$, and Jensen's inequality proves
Eq.~\eqref{eq:accumulated-trajectory-bound}.
The same argument bounds the mean and rms state-vector error with the
factor $2\norm{M}$ omitted.  No design assumption or independence between
successive trajectory states is needed.

It remains to insert the single-step product-formula estimate and the assumed
frame-potential upper bound.  Write $H=\sum_{\gamma=1}^{L}H_\gamma$ and define
the standard commutator scale
\begin{equation}
\widetilde\alpha_{\rm comm}:=
\sum_{\gamma_1,\ldots,\gamma_{p+1}=1}^{L}
\norm{[H_{\gamma_{p+1}},\ldots,[H_{\gamma_2},H_{\gamma_1}]\ldots]}.
\label{eq:alpha-comm}
\end{equation}
The one-step error of a $p$th-order formula satisfies
$\norm{D}=\mathcal O(\widetilde\alpha_{\rm comm}(\delta t)^{p+1})$
\cite{childs2021theory}.  To separate the one-step error scale from the
spectrum of $Q$, take a uniform upper bound $\chi_*$ on
$\chi_{\delta t}=\sqrt{\kappa_{\delta t}}/\norm{D}$ for $D\ne0$ over
the step sizes considered.
Such a bound exists at fixed $d$: the inequalities
$\Tr Q\le d\norm{D}^2$ and $\Tr(Q^2)\le d\norm{D}^4$ imply
$\chi_{\delta t}\le[d(d+1)/2]^{1/4}$.
If $D=0$, the simulation is exact.
By the frame-potential assumption in Proposition~\ref{thm:trajectory-trotter-scaling},
$F_2(T)\le C/T$ for a $T$-independent constant $C$ over the time range
considered.  
By definition,
$\sqrt{\kappa_{\delta t}}=\chi_{\delta t}\norm D\le\chi_*\norm D$.
Using the frame-dependent term of
Eq.~\eqref{eq:accumulated-trajectory-bound}, then substituting the one-step
estimate and $\delta t=T/r$, gives
\begin{equation}
\begin{aligned}
\epsilon_{{\rm rms},\delta t}
&\le2\norm M\,r\chi_*\norm D\left(\frac CT\right)^{1/4}\\
&\le\frac{2c_p\widetilde\alpha_{\rm comm}\norm M}{r^p}
\chi_*C^{1/4}T^{p+3/4}.
\end{aligned}
\label{eq:trotter-frame-step-bound}
\end{equation}
The time exponent follows from
$r(T/r)^{p+1}T^{-1/4}=T^{p+3/4}/r^p$.
Making the right-hand side at most $\epsilon$ requires
$r^p\ge(2c_p\widetilde\alpha_{\rm comm}\norm M\chi_*C^{1/4}/\epsilon)
T^{p+3/4}$.
Taking the $p$th root gives the exponent
$(p+3/4)/p=1+3/(4p)$ in
Eq.~\eqref{eq:trajectory-assisted-trotter-number}.
\begin{equation}
\epsilon_{{\rm rms},\delta t}
=O\!\left(\frac{\widetilde\alpha_{\rm comm}\norm{M}}{r^p}
\chi_* C^{1/4}T^{p+3/4}\right).
\label{eq:trotter-frame-step-bound-asymptotic}
\end{equation}
Solving for $r$ gives the number of Trotter steps sufficient to reach error
$\epsilon$, as stated in Eq.~\eqref{eq:trajectory-assisted-trotter-number}.
At fixed system size, the $T$-independent factor $(\chi_*C^{1/4})^{1/p}$
is absorbed into its hidden constant.  This proves Proposition~\ref{thm:trajectory-trotter-scaling}
within the stated time window and the small-step regime of the single-step bound.
\hfill$\square$

\textit{A fixed initial state.---}
For any fixed $s$, use the discrete ensemble
$\cE_{r,s}=\{r^{-1},\ket{\psi(s+j\delta t)}:j=0,\ldots,r-1\}$.
The same Cauchy--Schwarz argument, now without an offset integral, bounds
$e_r(s;T)\le2\norm{M}r\min\{\norm{D},
\sqrt{\kappa_{\delta t}}[F_2^{(r)}(T)]^{1/4}\}$, where
\begin{equation}
F_2^{(r)}(T)
=\frac1r+\frac{2}{r^2}\sum_{k=1}^{r-1}(r-k)C_2(k\delta t).
\label{eq:discrete-trajectory-frame}
\end{equation}
Time-translation invariance makes this scalar independent of $s$, including
$s=0$.  It remains a state-dependent bound because $C_2$ depends on the
specified $\ket{\psi_0}$.

To compare the discrete and continuous frames, expand
$C_2(\tau)=\sum_\ell w_\ell e^{-i\omega_\ell\tau}$.
Each weight is a product of four nonnegative spectral probabilities, so
$w_\ell\ge0$, and $|\omega_\ell|\le2W$ with $W=E_{\max}-E_{\min}$.
The continuous contribution of a frequency is
$w_\ell\operatorname{sinc}^2(\omega_\ell T/2)$; the discrete contribution
is this quantity divided by $\operatorname{sinc}^2(\omega_\ell \delta t/2)$.
For $\delta t W<\pi$, positivity and the monotonic decrease of sinc on $[0,\pi)$
imply
\begin{equation}
F_2(T)\le F_2^{(r)}(T)\le c_{\delta t}F_2(T),
\qquad c_{\delta t}=\operatorname{sinc}(\delta t W)^{-2}.
\label{eq:trajectory-frame-sampling}
\end{equation}
The expressions at zero frequency are understood by continuity.
This proves the fixed-state claim; $\delta t W\le\pi/2$ gives $c_{\delta t}\le\pi^2/4$.

\textit{Mean-square single-step error.---}
A refinement shows why a decaying frame alone does not guarantee an advantage.
Set $Q^0=Q-\norm{D}_F^2\I$, then $\Tr(Q^0)=0$ and $\Tr(Q^0\bar\rho_T)=\Tr[\mathcal A_{Q^0}M_2(\cE_T)]=\Tr(\mathcal A_{Q^0}\Delta_2)$, where $\Delta_2=M_2(\cE_T)-\frac{\I+\Swap}{d(d+1)}$. 
Set $\nu_{\delta t}=\sqrt{d(d+2)}\norm{Q^0}_F/2$.
Since
$d^2\norm{\Delta_2}_F^2=R_2(T):=F_2(T)-2/[d(d+1)]$,\\
\begin{equation}
\begin{aligned}
\Tr(Q\bar\rho_T)&=\norm{D}_F^2+\Tr(Q^0\bar\rho_T)\\
&=\norm{D}_F^2+\Tr(\mathcal A_{Q^0}\Delta_2)\\
&\le\norm{D}_F^2+\sqrt{\Tr(\mathcal A_{Q^0}^2)}\sqrt{\Tr(\Delta_2^2)}\\
&=\norm{D}_F^2+\nu_{\delta t}\sqrt{R_2(T)}.
\label{eq:local-defect-baseline}
\end{aligned}
\end{equation}
This supplies a third term
$\sqrt{\norm{D}_F^2+\nu_{\delta t}\sqrt{R_2(T)}}$ in the minimum in
Eq.~\eqref{eq:accumulated-trajectory-bound}.
For fixed $D$, only the second term on the right-hand side of
Eq.~\eqref{eq:local-defect-baseline} depends on the trajectory's frame potential.
Once $\nu_{\delta t}\sqrt{R_2(T)}\ll\norm{D}_F^2$, further reduction of
the excess frame potential gives only a small decrease in this single-step
upper bound.  The baseline does not impose a lower bound on the actual
accumulated observable error.

\textit{Number of Trotter steps and applicable time range.---}
The conventional bound follows by using only $\norm{D}$ in
Eq.~\eqref{eq:accumulated-trajectory-bound}, and gives
$r=O[(\widetilde\alpha_{\rm comm}\norm{M}/\epsilon)^{1/p}T^{1+1/p}]$.
For the frame-potential-assisted bound to improve upon the conventional operator-norm estimate, we require $\chi_{\delta t}F_2(T)^{1/4}<1$. Since $Q=D^\dagger D\succeq0$ and $\lambda_{\max}(Q)=\|D\|^2$, we have $\Tr(Q^2)\geq\|D\|^4$ and $\Tr(Q)\geq\|D\|^2$. Consequently,
\begin{equation}
\chi_{\delta t}^4
=\frac{(d+2)\Tr(Q^2)+[\Tr(Q)]^2}{4\|D\|^4}
\geq\frac{d+3}{4}.
\end{equation}
Thus, $F_2(T)<4/(d+3)$ is a necessary condition for an improvement.

The improved step-count exponent $1+3/(4p)$ applies at fixed system size
over the finite intermediate time range in which $F_2(T)=\mathcal O(T^{-1})$.
This regime ends when the frame-potential plateau dominates.
At a fixed step size, the baseline $\norm{D}_F^2$ can limit the improvement
due to frame-potential decay in Eq.~\eqref{eq:local-defect-baseline}
even before that plateau is reached.
In particular, the dimension and single-step error prefactors must be kept if the system
size changes with $T$.  If $H$ and $S_p(\delta t)$ preserve the same projector $P$
containing the initial state, the proof applies within $P\cH$ with
$d=\Tr P$, $D$ restricted to the sector, and $M$ replaced by $PMP$.

\textit{Relation to one-step observable errors.---}
One can also expand the accumulated error directly in the Heisenberg picture.  With
$M_j=([S_p(\delta t)]^{r-1-j})^\dagger M[S_p(\delta t)]^{r-1-j}$ and
$B_{\delta t,j}=U_0^\dagger(\delta t)M_jU_0(\delta t)
-S_p^\dagger(\delta t)M_jS_p(\delta t)$, the exact signed error is
\begin{equation}
\bra{\psi(s)}B_r(M;T)\ket{\psi(s)}
=\sum_{j=0}^{r-1}\bra{\psi(s+j\delta t)}B_{\delta t,j}\ket{\psi(s+j\delta t)},
\label{eq:observable-trajectory-telescoping}
\end{equation}
where $B_r(M;T)=U_0^\dagger(T)MU_0(T)-U_r^\dagger(T)MU_r(T)$.
The triangle inequality is valid for these evolved observables $M_j$; replacing
every $M_j$ by the fixed $M$ in Fig.~\ref{fig:single-trajectory-frame-error}
is not justified in general.  The state-vector error proof avoids that replacement
at the cost of retaining $\norm{D}_F^2$.  Testing the resource implication requires
evaluating the accumulated error under $U_r(T)$, the actual moments of $Q$, and the relevant
frame regime.  The current figure displays one-step ingredients only.

\section{Numerical settings}
\label{app:numerics}
All numerical figures use $H=H_x+H_y+H_{xx}$, where
$H_x=h_x\sum_{j=1}^nX_j$, $H_y=h_y\sum_{j=1}^nY_j$, and
$H_{xx}=J\sum_{j=1}^{n-1}X_jX_{j+1}$, with $h_x=0.809$, $h_y=0.9045$,
$J=1$.  The measured observable is
$M=(n-1)^{-1/2}\sum_{j=1}^{n-1}Z_jZ_{j+1}$.
Figures~1--3 use $\delta t=0.1$, the exact unitary
$U_0(\delta t)=\exp(-i\delta t H)$, and the first-order product-formula step
\begin{equation}
    S_1(\delta t)=\exp[-i\delta t(H_x+H_{xx})]\exp(-i\delta t H_y).
\end{equation}

\textit{Figure 1.---}
For each $4\le n\le8$, each curve uses $S=320$ product-Pauli or normalized
complex-Gaussian Haar states.  Each marker estimates
$S^{-1}\sum_{s=1}^{S}|X_{\psi_s}(M)|$; its bar is $s_{\rm abs}/\sqrt S$, where
$s_{\rm abs}$ is the sample standard deviation.  The deterministic 1- and
2-design bounds are $\norm{B(M)}_1/d$ and $\norm{B(M)}_F/\sqrt{d+1}$; the
latter is also the exact Haar root-mean-square error.

\textit{Figure 2.---}
For $n=10$, we generate two independent batches of $S_a=S_b=160$ samples.
Each sample consists of an independently sampled product-Pauli input and
an independently sampled brickwork circuit.  Every gate is drawn independently and
uniformly from the two-qubit Clifford group.  For each sample, we record
the input at depth $D=0$ and the state after each successive layer up to
$D=20$, using the same circuit as its depth increases.
Depth $D$ counts individual layers of disjoint nearest-neighbor gates.
Odd and even layers act on the bonds $(1,2),(3,4),\ldots$ and
$(2,3),(4,5),\ldots$, respectively.
At each depth, the observable curve uses the union of the two batches,
while the unbiased cross-batch estimator
$\widehat F_2(D)=(S_aS_b)^{-1}\sum_{a,b}
|\langle\psi_{a,D}^{(1)}|\psi_{b,D}^{(2)}\rangle|^4$ avoids finite-sample
self overlaps.  The right axis shows
$|\widehat F_2(D)-2/[d(d+1)]|^{1/2}$.  The horizontal mean-absolute Haar
reference uses 480 independently sampled states.

\textit{Figure 3.---}
For each initial state, exact diagonalization generates 64 snapshots at
$T=0.1,\ldots,6.4$.  At every snapshot, both branches defining $e(T)$ start
from the same exact state; the measurement time is $T+\delta t$.
Equation~\eqref{eq:trajectory-frame-potential} is evaluated
by composite Simpson quadrature with a uniform time spacing of $0.1/32$.
Halving this spacing in the zero-state calculation changes $F_2(T)$ by at most
$2.8\times10^{-8}$.

The initial-state energy variance is
$\sigma_E^2=\bra{\psi_0}H^2\ket{\psi_0}-\bra{\psi_0}H\ket{\psi_0}^2$.
For the zero (plus) state, $\sigma_E^2=23.7260$ ($8.18120$),
$D_{\rm eff}=238.125$ ($1.89009$), and
$C_2(\infty)=3.50\times10^{-5}$ ($0.295697$), with otherwise identical settings.
The gray dotted line in Fig.~\ref{fig:single-trajectory-frame-error} is the
exact Haar rms error $\norm{B(M)}_F/\sqrt{d+1}=1.48615\times10^{-3}$,
evaluated in the full $d=2^{10}=1024$ Hilbert space.
Here $\norm{B(M)}_F^2=\Tr[B(M)^2]/d$ and $\Tr B(M)=0$.
This baseline is shared by both panels and does not assume that either
trajectory samples Haar-random states.

\textit{Figure~\ref{fig:single-trajectory-pf2-resources}.---}
For $n=10$, we use
$S_2(\delta t)=e^{-iA\delta t/2}e^{-iB\delta t}e^{-iA\delta t/2}$,
with $A=H_x+H_{xx}$, $B=H_y$, and $\delta t=T/r$.
The error is $e_r(0;T)$, with the fixed input states and
$T=0.8,1.2,\ldots,12.8$.
For the plotted tolerance $\epsilon=0.01$, $r$ is the smallest scanned step
count whose error and those of all larger scanned counts are at most
$\epsilon$; it need not be the smallest passing count among all integers.
If the smallest passing count lies at the lower end of the scan, we extend
the scan to smaller step counts until a count with error above the tolerance
is found or $r=1$ is reached.
Fits minimize squared residuals in $\log r$ against $\log T$.
The reference is $r_{\rm ref}(T)=8(T/1.6)^{11/8}$.

\end{document}